\documentclass[a4paper]{jpconf}
\usepackage{graphicx}
\usepackage{xcolor}
\usepackage{amssymb}
\usepackage{amsmath}

\newcommand{\ND}{N_{\rm D}}
\newcommand{\NF}{N_{\rm F}}
\newcommand{\VD}{V_{\rm D}}

\newcommand{\rhoG}{\rho_{\rm g}}
\newcommand{\rhoL}{\rho_{\rm l}}
\newcommand{\rhoC}{\rho_{\rm c}}
\newcommand{\tauW}{\tau_{\rm W}}
\newcommand{\DeltaC}{\Delta_{\rm c}}
\newcommand{\lambdaC}{\lambda_{\rm c}}

\newcommand{\mykappa}{\hat{\kappa}}

\begin{document}
\title{From particle condensation to polymer aggregation}

\author{Wolfhard Janke$^1$ and Johannes Zierenberg$^{1,2,3}$}
\address{$^1$ Institut f\"ur Theoretische Physik, Universit\"at Leipzig, Postfach 100\,920, 04009 Leipzig, Germany }
\address{$^2$ Max Planck Institute for Dynamics and Self-Organization, Am Fassberg 17, 37077 G{\"o}ttingen, Germany}
\address{$^3$ Bernstein Center for Computational Neuroscience, Am Fassberg 17, 37077 G{\"o}ttingen, Germany}
\ead{wolfhard.janke@itp.uni-leipzig.de}

\begin{abstract}
We draw an analogy between droplet formation in dilute particle and polymer
systems. Our arguments are based on finite-size scaling results from studies of
a two-dimensional lattice gas to three-dimensional bead-spring polymers. To set
the results in perspective, we compare with in part rigorous theoretical scaling
laws for canonical condensation in a supersaturated gas at fixed
temperature, and derive corresponding scaling predictions for an undercooled gas
at fixed density. The latter allows one to efficiently employ parallel
multicanonical simulations and to reach previously not accessible scaling
regimes. While the asymptotic scaling can not be observed for the comparably
small polymer system sizes, they demonstrate an intermediate scaling regime also
observable for particle condensation. Altogether, 
our
extensive
results from computer simulations provide clear evidence for the close analogy
between particle condensation and polymer aggregation in dilute systems.
\end{abstract}

\section{Introduction}
Condensation and nucleation processes are underlying many phenomena from
galaxy evolution over cloud formation to protein aggregation. For particles 
or colloids much is known from experiments, theory and computation
\cite{kelton-frenkel-intro-nucl16, binder-virnau16, palberg16}. For polymers,
the situation is less settled.
At first sight, particle condensation and polymer aggregation appear to be very
different processes. While condensation is a phenomenon of one's
every-day-life and the stereotypic example of a first-order phase transition,
aggregation is usually associated 
with
neurodegenerative diseases such as
Altzheimer's, Parkinson's, and diabetes II~\cite{chiti2006}. However, if treated
outside of complex environments such as the brain, proteins in dilute solution
do show cluster formation~\cite{stradner2004} with a close resemblance to
classical nucleation theory~\cite{feder1966, oxtoby1992, kashchiev2000}.
Consequently, concepts such as nucleation rates and barriers can be
qualitatively carried over to more complex macromolecular
solutions~\cite{sear2007}.

There exist a lot more qualitative similarities between particle and polymer
systems. For example, the phase 
diagrams
of particle and polymer solutions
appear
quite similar~\cite{sear2007,wilding1995,wilding1997,macdowell2002,virnau2004,virnau2004b}. In
fact, already a single polymer shows similarities to a Lennard-Jones particle
cluster in its low-temperature structure~\cite{schnabel2009a,schnabel2009b,schnabel2011}.
Also, the multi-step nucleation of a few constituents to a full cluster shows a
comparable hierarchy by multiple backbending in a microcanonical analysis
depending on the
density~\cite{junghans2006,junghans2008,junghans2009,junghans2011,schierz2015,koci2017}.
For an abstract protein model as unit-length stick on a periodic cubic lattice
the formation of fibrils shows comparable scaling behavior as droplet formation,
which is expected since the model resembles a lattice gas with asymmetric
interactions~\cite{irbaeck2015}.  A natural question to ask is thus whether the
cluster formation in polymer systems shows the same scaling behavior as in
particle systems, despite their more complex structural
properties~\cite{zierenberg2016, janke2017}.

In this work, we want to recapitulate some evidence for an analogy between
cluster formation in a particle gas and a polymer solution using a consistent
model framework based on our own contributions in this field. By exploiting the
similarity with particle systems, we will analyze 
the implications for the finite-size scaling behavior in dilute polymer systems
and discuss how small system sizes may easily lead to misinterpretations.

\begin{figure}
  \centering
  \includegraphics[width=0.43\textwidth]{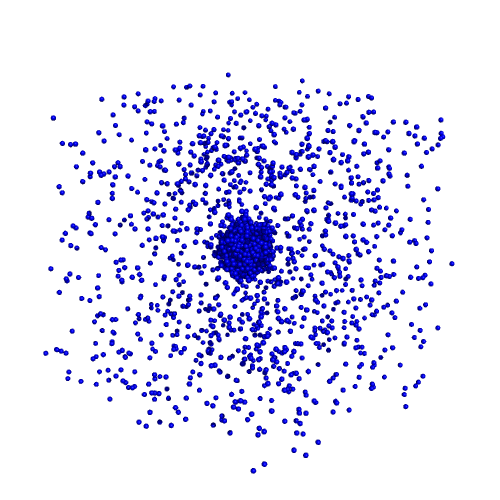}
  \hspace{2em}
  \includegraphics[width=0.45\textwidth]{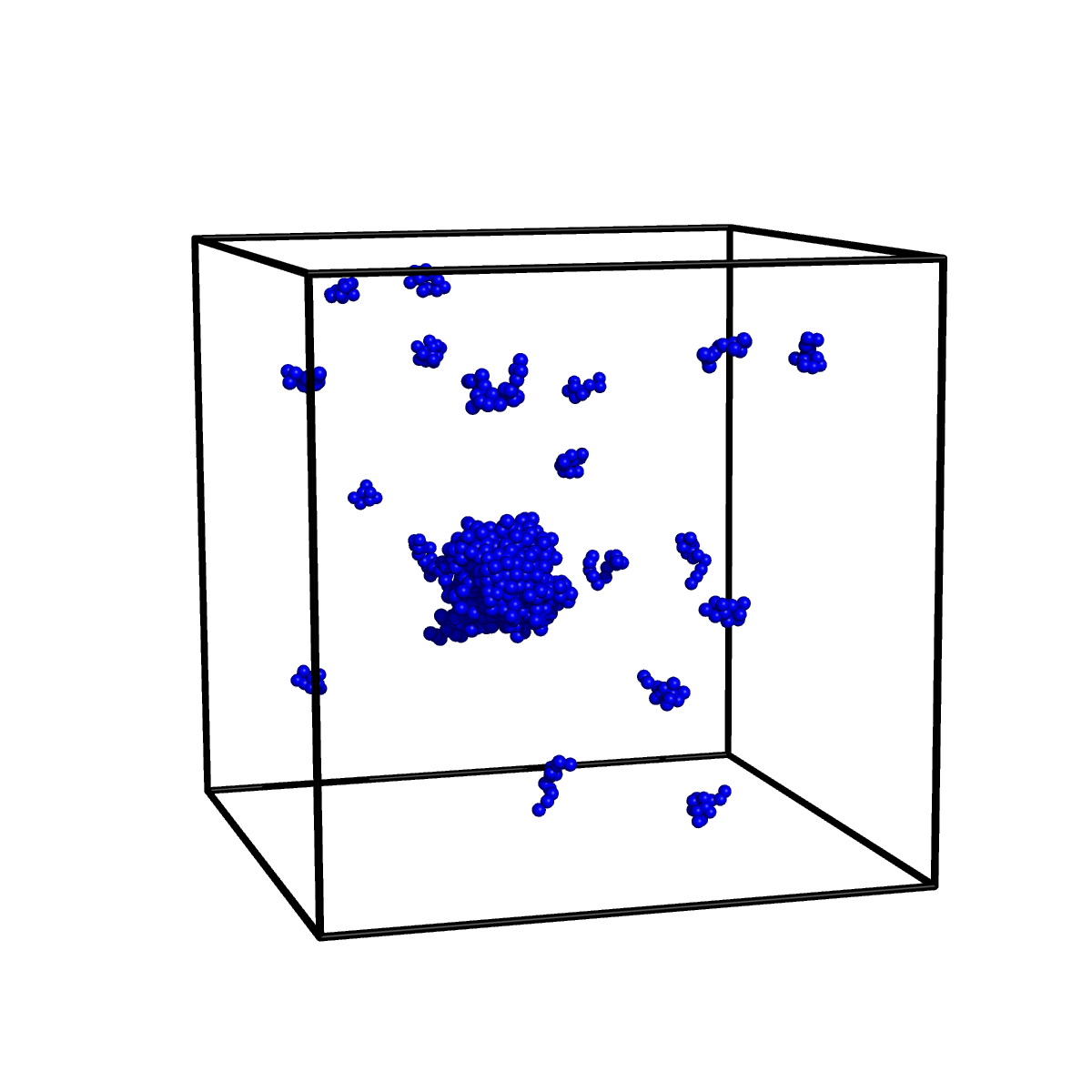}
  \caption{%
    Illustration of droplet formation in a system of $N=2000$
    particles (left) 
    and $N=64$ polymers of length
    $P=13$ (right).
    \label{figIllustration}
  }
\end{figure}

\section{Droplet formation in a particle gas}
\label{secParticles}

\subsection{General arguments involving competition of energy and entropy at
fixed temperature}

Let us consider the situation of a supersaturated gas in a finite system size
at fixed temperature, i.e., in the canonical ($NVT$)
ensemble~\cite{binder1980,biskup2002,biskup2003,neuhaus2003,binder2003}.
Supersaturation refers to a particle excess $\delta N = N-N_0$ compared to the
temperature-specific (grand-canonical) background gas density
$\rhoG(T)=N_0(T)/V$. Besides the constant free-energy contribution from the
background density, the particle excess can contribute to the system's free
energy either by an entropic part from the particle fluctuations $F_{\rm
fluc}$, or by an energetic part $F_{\rm drop}$ related to the formation of a
single macroscopic droplet of size $\VD$. For an illustration see
Fig.~\ref{figIllustration}~(left). The fluctuations can be treated in a
Gaussian approximation around the equilibrium background contribution and the
condensate is treated as an ideally shaped droplet \cite{biskup2002,
biskup2003}:
\begin{equation}
  F_{\rm fluc} = \frac{(\delta N)^2}{2\mykappa V}
  \qquad\text{and}\qquad
  F_{\rm drop} = \tauW \VD^\frac{d-1}{d}\,,
\end{equation}
where 
$\mykappa = \beta \kappa = \beta \big \langle (N - \langle N \rangle )^2 \big \rangle/V$
is the isothermal compressibility and $\tauW$ the surface free energy of a
(Wulff shaped) droplet of unit volume.

This treatment differs from the general view on nucleation free-energy barriers,
where interface free-energy loss competes with bulk free-energy gain because an
infinite reservoir of attaching particles is assumed. This essentially focuses
only on the droplet contribution $\Delta F_{\rm drop} = \VD\Delta f +
\partial\VD\sigma$, where $\Delta f$ is the free-energy difference per unit
volume, $\partial\VD$ denotes the surface of $\VD$, and $\sigma$ is the surface 
tension. This has a clear maximum, which is
obtained by solving $d \Delta F_{\rm drop}/dR=0$ for the critical droplet 
radius $R_c$. If $R>R_c$
the droplet will grow infinitely due to the infinite reservoir of particles. In
this case, $R_c$ is not system size dependent, such that all critical droplets
have the same size. We, instead, consider the equilibrium situation of a finite
particle number where the gas contribution needs to be considered explicitly.
For a good overview of the historical development, see, e.g.,
Ref.~\cite{schrader2009}.

\begin{figure}
\centering
\includegraphics[width=0.49\textwidth]{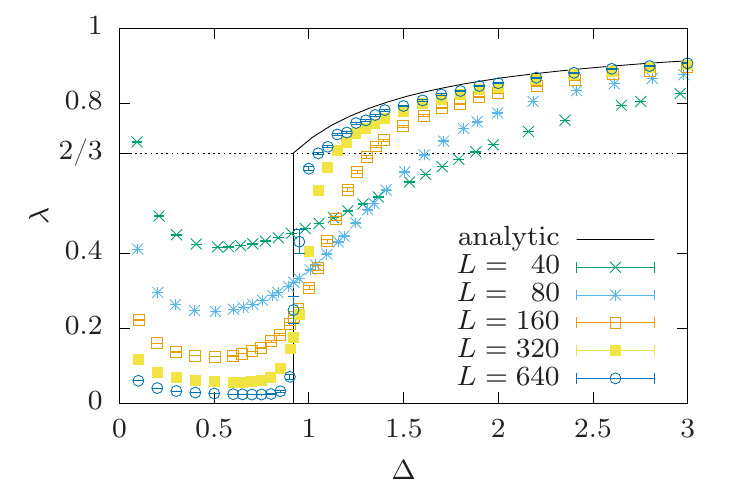}
\includegraphics[width=0.49\textwidth]{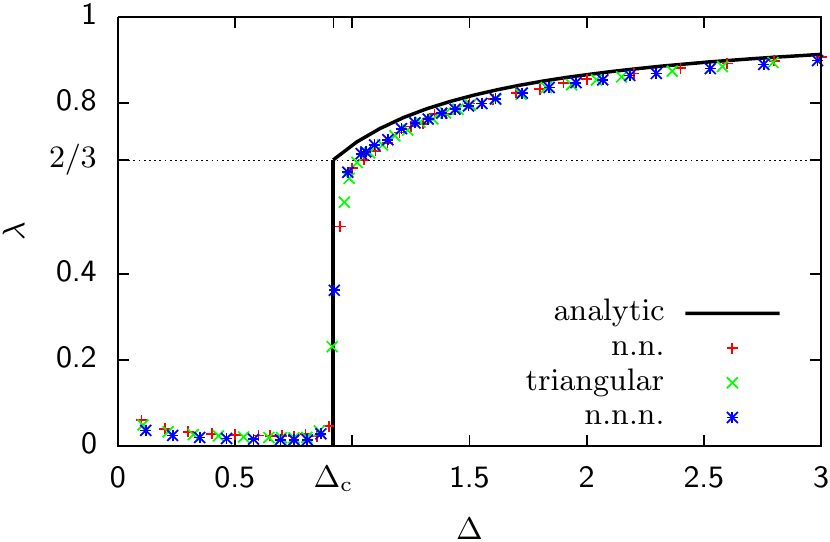}
\caption{%
  (left) Fraction of excess $\lambda$ that goes into the droplet as a function of
  dimensionless rescaled density $\Delta$ for 2D droplet formation in the
  Ising lattice gas (square lattice with nearest-neighbor interactions) at 
  fixed temperature $k_\mathrm{B}T=0.375$ (in lattice 
  gas normalization). Note that $\Delta$ already accounts for the
  leading-order finite-size scaling behavior of the density.
  (right) Comparison of results for three different 2D lattice gas models 
  (nearest-neighbor (n.n.) square lattice, n.n.\ triangular lattice, next-nearest-neighbor 
  (n.n.n.) square lattice) at $T \approx (2/3) T_c$ for large system size $L = 640$.
  Adapted from 
  Refs.~\cite{nussbaumer2006,nussbaumer2008,nussbaumer2010,nussbaumer2016}.
  \label{figGasFixT}
}
\end{figure}

We proceed our line of arguments by linking  the droplet size to the particle
excess inside the droplet, \mbox{$\delta\ND=(\rhoL -\rhoG)\VD$}, where $\rhoL$
and $\rhoG$ are the background liquid and gas density, respectively. General
arguments show that the probability of additional intermediate-sized droplets
vanishes~\cite{biskup2002, biskup2003}, so that we can distribute the particle
excess in a two-phase scenario between the excess inside the droplet $\delta
\ND$ and the excess in the fluctuating phase $\delta\NF$. In terms of the
scalar fraction
\begin{equation}
  \lambda = \delta\ND/\delta N\,,
  \label{eqLambda}
\end{equation}
we obtain $\delta\ND=\lambda \delta N$ and $\delta\NF = (1-\lambda)\delta N$.
Then, the total excess free energy $F=F_{\rm drop}+F_{\rm fluc}$ becomes
\begin{equation}
  F = \tauW \left( \frac{\lambda\delta N}{\rhoL-\rhoG}\right)^{\frac{d-1}{d}} + \frac{(1-\lambda)^2(\delta N)^2}{2\mykappa V}
    = \tau_W \left(\frac{\delta N}{\rhoL-\rhoG}\right)^{\frac{d-1}{d}}
       \left(\lambda^{\frac{d-1}{d}} + \Delta (1-\lambda)^2\right)\,,
  \label{eqCondensationFixTFreeEnergy}
\end{equation}
with a dimensionless ``density'' parameter
\begin{equation}
  \Delta
  = \frac{(\rhoL-\rhoG)^{\frac{d-1}{d}}}{2\mykappa \tau_W} \frac{(\delta
  N)^{\frac{d+1}{d}}}{V}
  = \frac{(\rhoL-\rhoG)^{\frac{d-1}{d}}}{2\mykappa \tauW} \left(\rho -
  \rhoG\right)^{\frac{d+1}{d}}~V^{\frac{1}{d}}\,.
  \label{eqCondensationFixTDelta}
\end{equation}
At fixed temperature, $\rhoL,\rhoG,\mykappa$, and $\tau_W$ are constants. By 
minimizing
Eq.~(\ref{eqCondensationFixTFreeEnergy}) with respect to $\lambda$ 
one obtains
the fraction of particles inside the largest droplet $\lambda$ as a function of
dimensionless density $\Delta$, as shown for a
two-dimensional lattice gas in Fig.\ \ref{figGasFixT}
by the black ``analytic'' line jumping at $\DeltaC = 
(1/2)(3/2)^{3/2} \approx 0.9186$ from $\lambda=0$ to $\lambda = 2/3$.
As derived in Refs.~\cite{biskup2002,
biskup2003}, there exists in general a constant
$\DeltaC=\frac{1}{d}\left(\frac{d+1}{2}\right)^{\frac{d+1}{d}}$ below which no
droplet forms ($\lambda=0$) and above which a single macroscopic droplet exists
with non-trivial $\lambda>\lambdaC=\frac{2}{d+1}$. This threshold already
includes the leading-order finite-size corrections, as can be seen if one rewrites
Eq.~(\ref{eqCondensationFixTDelta}) in terms of the density 
\begin{equation}
  \rhoC = \rhoG +
  \left(\frac{2\mykappa\tauW\DeltaC}{(\rhoL-\rhoG)^{\frac{d-1}{d}}}\right)^{\frac{d}{d+1}}~V^{-\frac{1}{d+1}}\,.
  \label{eqCondensationFixTCorrections}
\end{equation}
It is less well visible that the critical droplet size $V_c=\lambda_c\delta
N_c/(\rhoL-\rhoG)$ scales non-trivially with system size, namely $V_c\propto
R_c^d\propto V^{d/(d+1)}$~\cite{binder1980,binder2003,zierenberg2015}.

Figure~\ref{figGasFixT} compares the results of simulations with the analytic
prediction for a (square) lattice gas with nearest-neighbor interactions in
2D~\cite{nussbaumer2006,nussbaumer2008,nussbaumer2010,nussbaumer2016}. 
In the lattice gas model, particles are modeled as non-empty lattice sites. 
This is equivalent to
an Ising model, where particles correspond to either up or down spins while the
vacuum corresponds to the other direction. For instance, identifying particles
with up spins, the particle density $\rho$ is related to the magnetization of
the Ising model by $\rho = N/V = (1+m)/2$ and the particle gas temperature 
is
$T = T^{\rm Ising}/4$.  The data in
Refs.~\cite{nussbaumer2006,nussbaumer2008,nussbaumer2010,nussbaumer2016} 
was generated in this
framework using mostly fixed-magnetization Metropolis~\cite{metropolis1953}
simulations (Kawasaki dynamics) at a given temperature $k_\mathrm{B}T = 0.375$.
In two dimensions, the equivalence to the Ising model additionally provides
exact results or very precise estimates for the infinite-size quantities
$\rhoG$, $\rhoL=1-\rhoG$, $\mykappa$, and $\tauW$, necessary for the rescaling
of the data according to Eq.~(\ref{eqCondensationFixTDelta}) in order to
compare with the theoretical (and, for the square lattice, rigorous
\cite{biskup2002, biskup2003})
prediction~\cite{nussbaumer2006,nussbaumer2008,nussbaumer2010,nussbaumer2016}. 

The data in Fig.~\ref{figGasFixT}~(left) clearly shows that for systems of 
finite size the analytic solution is overestimated for low densities and 
(slightly) underestimated for densities above the predicted transition density 
$\Delta_c$. The low-density finite-size effect is an unavoidable artefact due 
to the non-zero size of the largest cluster in the gas phase, which is at least
$\mathcal{O}(1/N)$. The high-density deviations, on the other hand, can be
attributed to physical effects such as non-ideal droplet shapes, including
capillary wave distortions, and further non-trivial corrections to the theory
\cite{kotecky-private}. As theoretically expected, these deviations
rapidly vanish in the limit of large system size. In 
Refs.~\cite{nussbaumer2008,nussbaumer2010} 
we have also checked that the
analytic solution, which strictly speaking was derived for the square lattice
with nearest-neighbor (n.n.) interactions in the infinite-size limit, describes 
other 2D lattice gases as well. This is demonstrated in 
Fig.~\ref{figGasFixT}~(right) where data for the next-neighbor triangular 
and next-nearest-neighbor (n.n.n.) square lattice gas are compared with
the standard n.n.\ square lattice reference case.

Overall, Fig.~\ref{figGasFixT} shows a good agreement
with a surprising symmetry around $\Delta_c$. This may be caused by a 
judicious
choice of the fixed temperature, far enough away from the critical point to
avoid strong fluctuations but
high enough to avoid slow dynamics. Results for $\lambda(\Delta)$ in 3D lattice
models are less symmetric, see, e.g., Ref.~\cite{zierenberg2014jpcs}, and are
currently under closer investigation. The leading-order phenomenological
scaling theory of droplet formation was in general found to be consistent with
numerical results for both 3D Lennard-Jones~\cite{macdowell2004} and
lattice~\cite{schmitz2013} models.
%

\subsection{Changing perspective to fixed density}
The theory developed for fixed temperature quantitatively describes the
formation and growth of a droplet with supersaturation, already including the
leading-order finite-size corrections. Now, we consider instead a particle 
system at
fixed density $\rho$ where the temperature is lowered until a droplet forms in an
\emph{undercooled} environment, again in the canonical ($NVT$) ensemble as in
Refs.~\cite{martinos2007,zierenberg2015,zierenberg2016athens,zierenberg2017NatComm}. This
can be seen as an ``orthogonal'' perspective in the temperature-density phase
diagram of the liquid-vapor transition~\cite{zierenberg2015}. Considering that the
``density'' parameter $\Delta$ already includes the dominant or leading-order finite-size
scaling behavior, we can 
rewrite Eq.~(\ref{eqCondensationFixTDelta}) as $V^{-1/(d+1)}\Delta^{d/(d+1)}=f(\rho,\beta)$,
which admits a power-series expansion in $\beta$ around the infinite-size inverse
temperature $\beta_0$ where $\rho_g(\beta_0) = \rho$. The solution of
$\Delta = \Delta_c$ then results in an asymptotic expansion in $V^{-1/(d+1)}$ for
the droplet-formation point $\beta(N)$ of a finite number of particles $N$.
Using the Taylor 
expansion 
$f(\rho,\beta) =
f(\rho,\beta_0)+f'(\rho,\beta_0)(\beta-\beta_0)+\frac{1}{2}f''(\rho,\beta_0)(\beta-\beta_0)^2+...$, 
where $f(\rho,\beta_0)=0$,
one obtains
%
\begin{equation}
	\beta(N) = \beta_0 + \frac{\Delta_c^{d/(d+1)}}{f'(\rho,\beta_0)} V^{-1/(d+1)} -
	\frac{f''(\rho,\beta_0)\Delta_c^{2d/(d+1)}}{2f'(\rho,\beta_0)^3} V^{-2/(d+1)}+
  \mathcal{O}\left(V^{-3/(d+1)}\right)\,.
\end{equation}
This of course neglects higher-order corrections to the fixed-temperature
solution, especially expected non-trivial and logarithmic
terms~\cite{kotecky-private,langer1967,ryu2010,nussbaumer2010,prestipino2012}.
However, the scaling corrections correspond to powers of the critical droplet
radius 
$R_c\propto V_D^{1/d}\propto V^{1/(d+1)}$~\cite{binder2003,zierenberg2015}
and we will show that fits to the higher-order terms from the Taylor expansion
nicely describe the finite-size corrections in the data. 
Since the density is fixed, we can further replace $V\propto N$ and arrive at the ansatz for
droplet formation in 3D
\begin{equation}
  \beta(N) = \beta_0 + a N^{-1/4} + b N^{-1/2} + c N^{-3/4}\,.
  \label{eqAnsatzBeta}
\end{equation}

\begin{figure}
\centering
\includegraphics[width=0.49\textwidth]{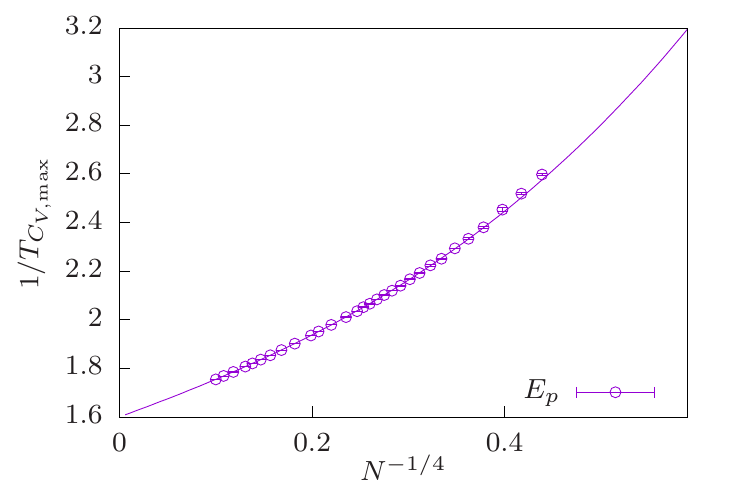}
\includegraphics[width=0.49\textwidth]{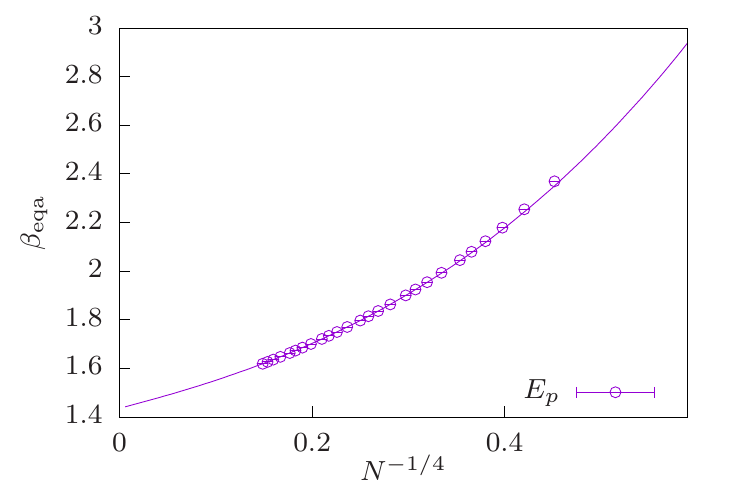}
\caption{%
  Finite-size scaling of the inverse transition temperature (fixed density
  $\rho=10^{-2}$) for 3D droplet formation in (left) a lattice gas and (right) a
  Lennard-Jones gas. For the lattice gas the density is of course
  approximate due to discretization effects. The solid lines show fits to
  the ansatz (\ref{eqAnsatzBeta}).
  Adapted from Refs.~\cite{zierenberg2015,zierenberg2017NatComm}.
  \label{figGasFigRho}
  }
\end{figure}
Figure~\ref{figGasFigRho} shows data for droplet formation in a 3D lattice and
Lennard-Jones gas based on data of Refs.~\cite{zierenberg2015,zierenberg2017NatComm}.
The lattice gas is directly treated as occupied sites on a cubic lattice with
Ising-like nearest-neighbor interactions, while
Lennard-Jones particles have a real-valued position in a cubic box with periodic
boundary conditions interacting via the potential 
$V_{\rm LJ} = 4\epsilon \big [(\sigma/r)^{12}-(\sigma/r)^{6} \big ]$ which 
exhibits a minimum at $r = 2^{1/6}\sigma$ of depth $-\epsilon$. Both data sets 
were generated using multicanonical
simulations~\cite{berg1991,berg1992,janke1992,janke1998}
in a parallelized implementation \cite{zierenberg2013,zierenberg2014pp},
which is a perfect choice for first-order-like phase transitions, especially 
upon variation of a continuously
variable control parameter such as the temperature (for recent reviews, see
Refs.~\cite{wj-paul-review2016, berg-review2017}). The parallel implementation 
shows linear speedup for particle condensation~\cite{zierenberg2014jpcs}. 
Canonical expectation values are obtained after a production run by standard 
reweighting techniques~\cite{janke2003}.
The finite-size transition temperatures $T_{C_V, {\rm max}}$ can be read off 
from the peak locations of the specific
heat $C_V=\beta^2\left(\langle E_p^2\rangle - \langle E_p \rangle^2\right)/N$ 
(where $E_p$ is the potential energy; see below) in
case of the lattice gas; and from an equal-area construction in the microcanonical
inverse temperature $\beta(E_p)$ \cite{janke-micro} 
in case of the Lennard-Jones gas, denoted in Fig.~\ref{figGasFigRho}~(right) 
by $\beta_{\rm eqa}$. Since we have
no analytical solution for the infinite-size quantities $\rho(T)$, $\kappa(T)$,
or $\tau_W(T)$, we merely fit the amplitudes of the correction terms in
Eq.~(\ref{eqAnsatzBeta}). In both
cases, the fit qualitatively describes the data very well. For the lattice gas,
the fit spans $N\in[106,10\,000]$ with a goodness-of-fit parameter $Q\approx0.002$,
attributed to the merely approximate density that one can adjust at each value of $N$. For the
Lennard-Jones gas, the fit spans $N\in[192,2048]$ with $Q\approx0.5$. 

Overall, the data is in good agreement with the theory on droplet formation in an
undercooled gas. It is worth noting that the comparably large finite-size
corrections may enable a systematic study of finite-size scaling in
experimental setups on the nanoscale. If one considers argon as a model system,
our largest 
numerical Lennard-Jones setup with $2048$ particles corresponds to
a box of size $L'\approx59~r'_{\rm min} = 59\times~2^{1/6}~\sigma'$ and with
$\sigma'\approx3.4\mathring{A}$ we find $L'\approx22.5~{\rm nm}$. This is a
scale that both experiments and simulations may reach.

\section{Droplet formation in a solution of flexible polymers}
Polymers are extended objects which bring an additional conformational entropy
if isolated in the high-temperature phase~\cite{deGennes}.
Lowering the temperature results in the formation of a macroscopic aggregate
surrounded by a gas of isolated
chains~\cite{zierenberg2017NatComm,zierenberg2014, mueller2015}, for an
illustration see Fig.~\ref{figIllustration}~(right). The shape of an aggregate
of flexible polymers, however, shows strong structural similarities to the
droplet in a gas~\cite{schnabel2009b,zierenberg2014}, with polymers actively
unfolding when incorporated into the cluster~\cite{zierenberg2014,mueller2015}.
If one neglects the single-polymer free-energy contributions, one may consider
polymers as extended particles with a more complex, potentially temperature
dependent, interaction potential. Then, it would make sense to expect the same
scaling behavior for the (inverse) transition temperature as predicted for
particle condensation and outlined in Sec.~\ref{secParticles}.

\begin{figure}
\centering
\includegraphics[width=0.49\textwidth]{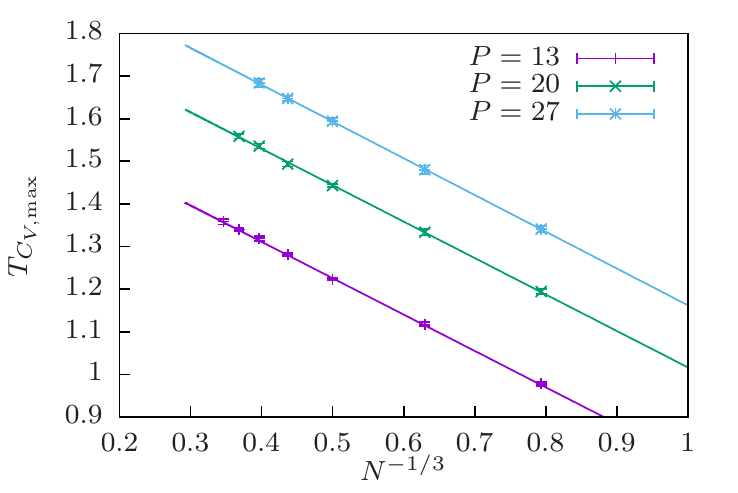}
\includegraphics[width=0.49\textwidth]{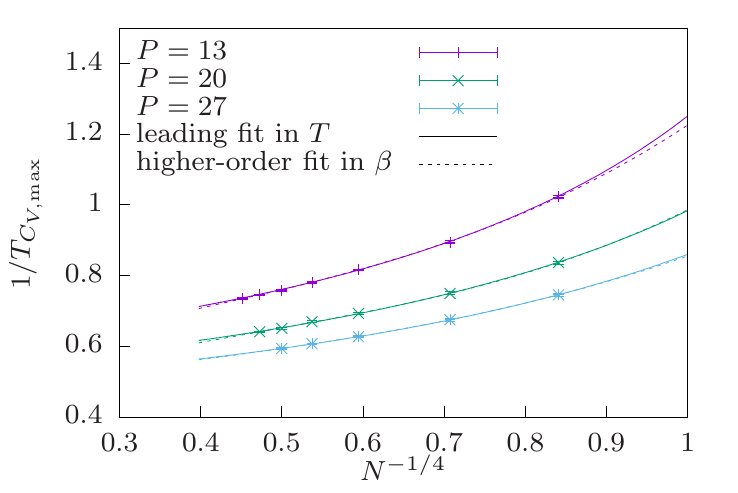}
\caption{%
  (left) Finite-size scaling of the transition temperature (fixed polymer
  density $N/V = 10^{-3}$) of cluster formation in a solution of bead-spring
  polymers of different length $P$ suggests a scaling as $T-T_0\propto N^{-1/3}$
  (solid lines). 
  (right) If the inverse temperature is instead fitted with higher-order terms
  the leading behavior $\propto N^{-1/4}$ is consistent with 
  Eq.~(\ref{eqAnsatzBeta}) (dashed lines). Adapted from Ref.~\cite{zierenberg2014}.
  \label{figPolyOff}
}
\end{figure}
When treating comparably small system sizes, we noticed, however, that the
scaling of the aggregation transition temperature for flexible bead-spring
polymers, obtained from parallel multicanonical simulations with up to 
$N = 24$
polymers of length $P=13$ 
(``13mers''), 20 20mers 
or 16 27mers,
shows an apparent
$T-T_0\propto N^{-1/3}$ scaling behavior~\cite{zierenberg2014}, see
Fig.~\ref{figPolyOff}~(left). This appears quite puzzling in the light of our
above arguments. However, inspired by the findings for polymers, a close
reexamination of particle condensation revealed a prominent \emph{intermediate
scaling regime}~\cite{zierenberg2015}, consistent with the $N^{-1/d}$ scaling
behavior observed for polymers in 3D~\cite{zierenberg2014}. Physically this
means that for a small number of constituents either none or {\em all\/} of
them assemble in the droplet -- the surrounding gas phase cannot really
develop. The problem is, that
even for particles this regime extends up to 
$N=10^3-10^4$, 
also depending on the dimension~\cite{zierenberg2015}. This is out of scope even of
modern simulation tools for equilibrium estimates of polymer aggregation. 

One may now argue that the observation of this intermediate regime is already a
first hint on an analogy between cluster formation in particle and polymer
systems. In addition, we observe that the empirical corrections from
Eq.~(\ref{eqAnsatzBeta}), as powers of the critical cluster size, describe the
scaling of the inverse transition temperature quite well
[Fig.~\ref{figPolyOff}~(right), dashed lines] and is almost indistinguishable
from the leading-order fit to the transition temperature
[Fig.~\ref{figPolyOff}~(right), solid lines]. 

\begin{figure}
\centering
\includegraphics[width=0.49\textwidth]{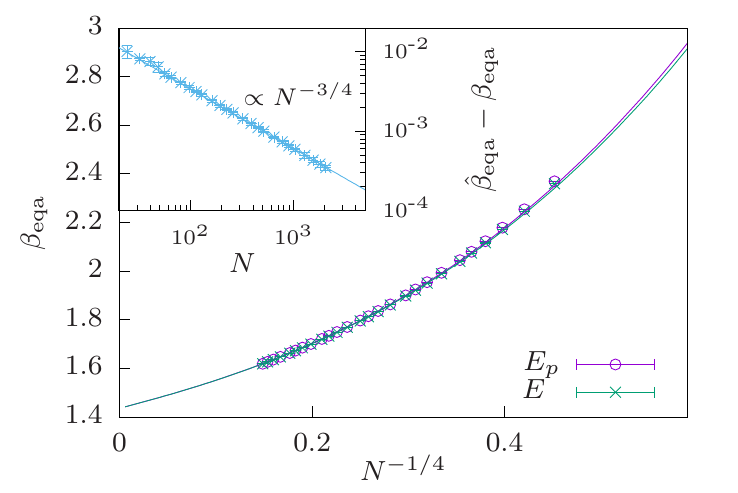}
\includegraphics[width=0.49\textwidth]{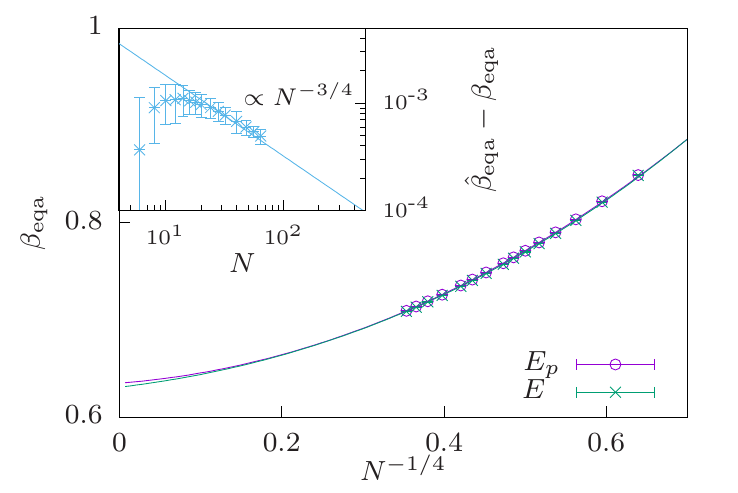}
\caption{%
  Finite-size scaling of the inverse transition temperatures (fixed monomer density
  $NP/V=10^{-2}$) for 3D cluster formation in (left) a gas of Lennard-Jones
  particles and (right) a solution of flexible bead-spring polymers 
  ($P = 13$ monomers each) reveals the same leading-order scaling as 
  $\beta\sim N^{-1/4}$.
  The transition point is determined by the equal-area rule using the
  microcanonical inverse temperature as a function of potential energy $E_p$ or
  total energy $E$. 
  Adapted from Ref.~\cite{zierenberg2017NatComm}.
  \label{figGasPoly}
}
\end{figure}

Treating larger polymer systems with up to $N=64$ polymers of length $P=13$
in Ref.~\cite{zierenberg2017NatComm}
further supports the strong similarity between particles and polymers,
again obtained from parallel
multicanonical simulations. The inverse temperatures plotted in 
Fig.~\ref{figGasPoly} are obtained from the
equal-area construction in the microcanonical inverse temperature as a function
of energy (note that a {\em monomer\/} density of $10^{-2}$ corresponds for
13mers to a {\em polymer\/} density of $\approx 10^{-3}$ as used in Fig.~\ref{figPolyOff}).
Here we also emphasize a conceptual point by using both
the potential energy [$\beta(E_p)$] as used is most work using
Monte Carlo simulations and the total energy [$\beta(E)$] as considered
in any textbook on statistical physics. In the latter case we
analytically included the kinetic energy $E_k$ what, technically, can be achieved
by a convolution of the potential-energy probability distribution $P(E_p)$ with
the Maxwell-Boltzmann distribution
$P(E_k)=\frac{\beta^{3N/2}}{\Gamma(3N/2)}E_k^{(3N-2)/2}e^{-\beta E_k}$.
For details see Ref.~\cite{zierenberg2017NatComm}. In Ref.~\cite{schierz2016} a
complementary simulation method is discussed that samples directly the
``real'' microcanonical ensemble including the kinetic energy 
contribution \cite{calvo2000,martinmayor2007}.
As expected both definitions
show the same behavior with very small differences. It is interesting, however,
that the difference between both definitions scales for large $N$ as
$\hat{\beta}-\beta\propto N^{-3/4}$, i.e., with the volume of the critical
cluster~\cite{zierenberg2015}. This systematic behavior can be anticipated also
for the case of a polymer solution. In total, this supports our hypothesis that
the formation of a cluster in a dilute solution of flexible polymers can, in 
leading order, be described by the generic arguments derived for particle systems.

\begin{figure}
\centering
\includegraphics[width=1.0\textwidth]{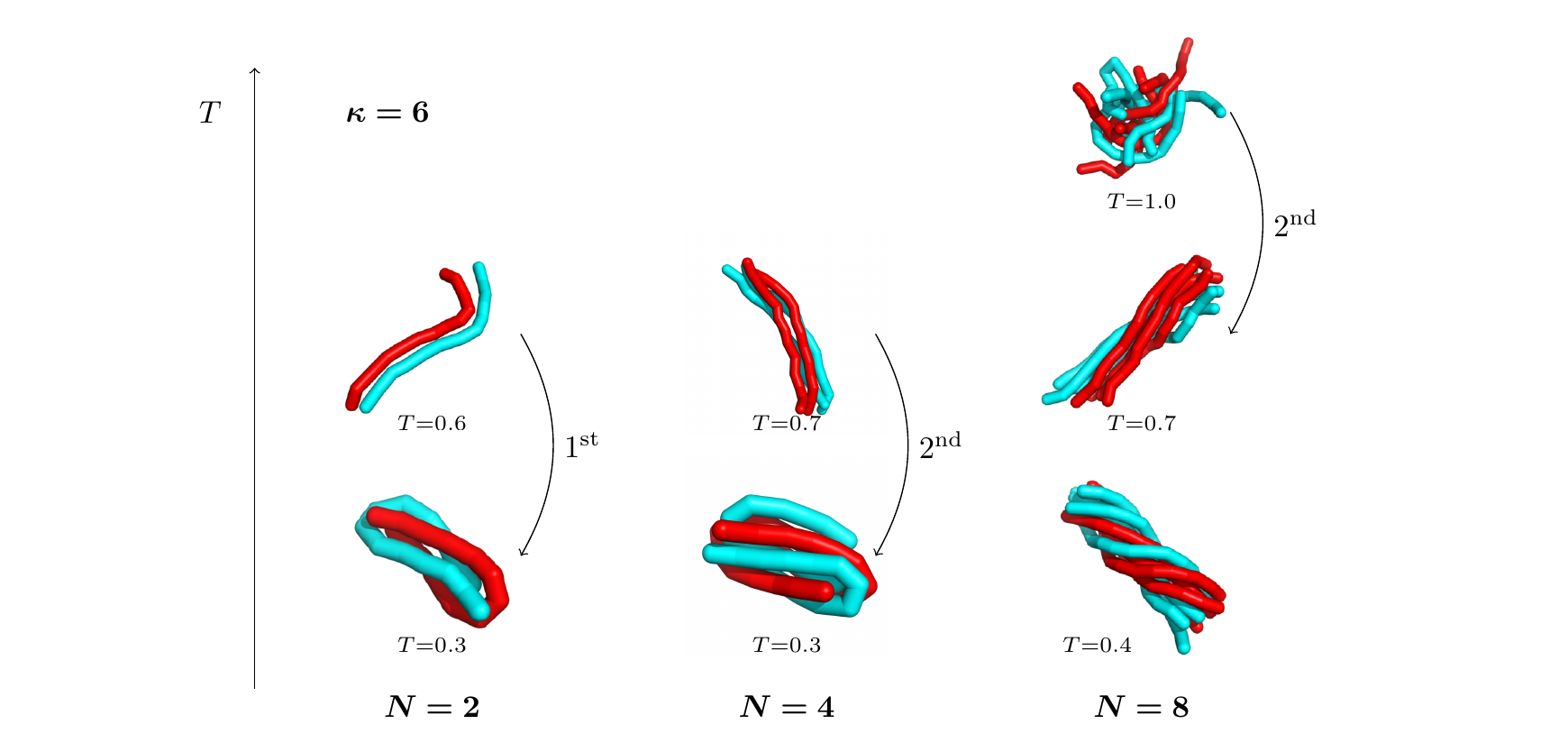}
\caption{%
Typical aggregate morphologies of $N=2,4$, and 8 semiflexible polymers 
of length $P=13$ in the intermediate-stiffness regime with bending-stiffness
constant $\kappa = 6$, illustrating the size-dependence of the sub-aggregation 
re-ordering transitions within the aggregate: $N=2$ polymers exhibit a 
first-order-like bundle-to-hairpin structural transition, whereas the 
finite-size transition for $N=4$ polymers from elongated bundles to bundled 
hairpins is second-order-like. $N=8$ polymers undergo a second-order-like 
amorphous-to-bundle transition followed by the formation of twisted bundles 
at even lower temperatures~\cite{zierenberg2015epl}.
\label{figSemiflexConfs}
}
\end{figure}

As an outlook to future work it should be stressed that biological systems 
usually are comprised of more complex macromolecules that cannot be
assumed to be flexible. A straightforward extension of the flexible
bead-spring polymer model considered above is to include worm-like chain 
motivated bending stiffness, leading to semiflexible homopolymers. This 
has an immediate impact on the structure of the low-temperature 
aggregate~\cite{zierenberg2015epl, zierenberg2016, janke2017}, 
cf.\ Fig.~\ref{figSemiflexConfs}. While rather flexible polymers keep on
forming spherical aggregates, sufficiently stiff polymers form (twisted)
bundles~\cite{kierfeld2005,heusinger2010,zierenberg2015epl,zierenberg2016, janke2017} 
known from biopolymer systems, e.g., from 
amyloid protofibrils \cite{giurleo-etal2008} or 
actin networks \cite{pandolfi-etal2014}, or 
polymeric materials \cite{kouwer-etal2013}. These bundles are not
spherical but for finite diameter resemble a cylindrical structure. We do
not suspect this to play an important role on the scaling of the cluster
formation transition as long as the shape of the cluster does not become
fractal. This is currently under active research.

\section{Conclusions}
\label{secConclusions}
In this paper we have compiled quite an extensive survey of computer
simulation studies of particle condensation and polymer aggregation.
By analyzing and interpreting the numerical data in a consistent
theoretical framework, 
clear evidence for a
close analogy between the cooperative behavior of particles and
flexible 
polymers 
emerges.

To arrive at this main conclusion, we started with a lattice gas
formulation at fixed temperature in 2D, for which we confirmed and
extended in part rigorous theoretical scaling predictions in the
infinite-size limit. We then proceeded to lattice and continuum
Lennard-Jones particle systems in 3D, for which an ``orthogonal''
constant density approach 
was demonstrated to be more efficient. This
enabled
studies of large enough particle numbers
to identify an intermediate scaling regime for 
smaller systems.
Taking this crossover behavior into account
we could finally
show
that the apparently more involved aggregation process of
flexible polymers is governed by the same
asymptotic
scaling laws as for particles.

As an outlook we briefly mentioned on the technical side computer
simulations in the ``real'' microcanonical ensemble that appear
competitive to generalized-ensemble methods and on the physics
side the more intricate aggregation behavior of semiflexible
polymers. 
In general, our approach can be extended to elucidate
nucleation mechanisms including nucleation barriers and
rates in greater detail.
These topics appear to be promising avenues 
for
future work.


\section*{Acknowledgments}
This work has been partially supported by
the Deutsche Forschungsgemeinschaft (DFG) under
Grant No.\ JA 483/31-1 and SFB/TRR 102 (project B04),
the Leipzig Graduate School of Natural Sciences ``BuildMoNa'',
the Deutsch-Franz\"osische Hochschule DFH-UFA through the Doctoral College 
``${\mathbb L}^4$'' under Grant No.\ CDFA-02-07,
and the EU Marie Curie IRSES network DIONICOS under Contract 
No.\ PIRSES-GA-2013-612707.
JZ received financial support from the German Ministry of Education
and Research (BMBF) via the Bernstein Center for Computational Neuroscience
(BCCN) G{\"o}ttingen under Grant No.~01GQ1005B.

\end{document}